\documentclass[preprint, 12pts]{aastex}
\usepackage[english]{babel}
\usepackage{subfigure,epsfig, geometry, times, amsmath}

\def\etal{et al.\ }

\def\pac{Paczy\'{n}ski~}
\def\woz{Wo\'{z}niak~}

\def\tmax{{t_{\rm max}}}
\def\fll{f_{\rm ll}}
\def\te{{t_E}}
\def\tep{{t_E^\prime}}
\def\Amax{A_{\rm max}}
\def\Amaxp{A^\prime_{\rm max}}
\def\umin{u_{\rm min}}
\def\uminp{u^\prime_{\rm min}}

\begin{document}

\title{Fitting Photometry of Blended Microlensing Events}

\author{      
      Christian L.~Thomas\altaffilmark{1}  and
      Kim Griest\altaffilmark{1}}

\altaffiltext{1}{Department of Physics, University of California,
    San Diego, CA 92093, USA\\
    Email: {\tt clt, kgriest@ucsd.edu }}

\begin{abstract} 
We reexamine the usefulness of fitting blended lightcurve models to
microlensing photometric data.  We find agreement with previous workers 
(e.g. \woz \& \pac) that this is a difficult proposition because 
of the degeneracy of blend fraction with other fit parameters. 
We show that follow-up observations at specific point along the lightcurve
(peak region and wings) of high magnification events are the most 
helpful in removing degeneracies.  
We also show that very small errors in the baseline
magnitude can result in problems in measuring the blend fraction,
and study the importance of non-Gaussian errors in the fit results.  The biases and skewness in the distribution of the recovered blend fraction is discussed.
We also find a new approximation formula relating the blend fraction
and the unblended fit parameters to the underlying event duration 
needed to estimate microlensing optical depth.

\end{abstract}


\section{Introduction}

Gravitational microlensing has become a useful tool in measuring the amount
of matter along the line-of-sight to distant stars.  
Since gravitational lensing depends only on mass, microlensing
is sensitive to all compact forms of matter independent of their luminosity.
Thus measurements out of the plane of the Galaxy towards the LMC, SMC,
and M31
have given important limits/detections of dark matter in the halo (
Aubourg, \etal\ 1993; Lasserre \etal\ 2000; Alcock, \etal 1993; 1997a; 2000;
Paulin-Henriksson \etal 2003; de Jong \etal 2004), 
and measurements towards the Galactic bulge give 
important constraints on the mass and distribution of Galactic stars, 
including those too faint to 
observe directly (Griest et al. 1991; \pac 1991; Han \& Gould 2003;
Udalski, et al. 1994; Alcock, et al. 1997b; Afonso, et al. 2003; 
Popowski, et al. 2005; Sumi, et al. 2005).   

The signal of microlensing is a specific transient magnification of 
a background source star as the lens object passes in front of it,
and thus microlensing experiments repeatedly monitor many ordinary stars to
find microlensing lightcurves.
The probability of microlensing occurring to a given star is called
the optical depth, $\tau$ and is of order $10^{-6}$ or less
for many Milky Way lines-of-sight.  The smallness of $\tau$ means that
microlensing experiments concentrate on very crowded star fields where
many hundreds of thousands of stars can be simultaneously imaged.  
This allows many lightcurves to be created simultaneously but also
results in blending of the source stars together.
This blending causes two problems in using the detected microlensing
events to infer the optical depth.  First, since each ``source object" may
contain the light from many stars, the number of stars being monitored is
not just the number of objects being photometered.  Second, the magnification profile of a microlensing event is changed when unlensed light is blended with the lensed light of the source star.

In this paper we revisit the problem of blending in microlensing 
lightcurves.  
There are several methods of dealing with the blending problem.
Among these are 1) obtaining high resolution images from space, 
which will usually allow separationof the source object 
into its different components, 
giving a direct measurement of the fraction of light from the lensed source (Alcock, et al. 2001),
2) if the unlensed light is not exactly centered on the lensed source,
then the centroid of the light will shift during the microlensing
event allowing limits on blending to be placed (Alard, Mao, \& Guibert 1995), 
3) if the lensed source 
is a different color than the unlensed light then 
a color shift will occur as the event proceeds, allowing
limits on lensed-light fraction to be made (Alard, Mao, \& Guibert 1995), and
4) for image subtraction lightcurves, the source can in principle be removed
and this can help break the degeneracy in some cases (Gould \& An 2002).

However, we will not discuss the above methods in this paper but will focus
on the fitting and interpretation of the photometric data alone; that
is, we include the lensed-light fraction as a parameter in the microlensing
fit and hope to use the shape of the lightcurve to recover this information. 
In principle this
allows recovery of the actual event duration, and a measurement
of the amount of blending in the sample of events, allowing corrections
to be made in estimating the optical depth.  A related and popular method
is to calculate lensing optical depth using only a subsample
of very bright source stars (e.g. clump giants).  The idea is
that very bright source stars are less likely to be blended, and when they
are blended, should be blended only by a small amount.
In this case one would like to use the blend fits only to determine
whether or not a given event is blended.

Unfortunately, as pointed out previously 
(e.g. Han 1999; DiStefano \& Esin 1995; \woz \& \pac 1997;
Alard 1997, etc.) blended fits
tend to be quite degenerate.  A lightcurve with a small lensed-light fraction
looks very much like an unblended lightcurve with a smaller maximum
magnification and a smaller event duration.  As pointed out
previously, this means that this fitting method
will be of limited use in many cases.  Our study adds strength to the
conclusions of previous workers, points out several new problems with
blend fits, and makes recommendations on how best to proceed with blend fits
for those who choose to do them.  We will discuss what happens 
when the microlensing
event contains signal from other physical effects such as weak parallax
or binary effects.  These effects are not rare, and since the difference
between blended and unblended lightcurves is small, even an almost
undetectable real deviation
from the standard point-source-point-lens lightcurve can render blend fit
results meaningless.

The plan of the paper is as follows:
In \S~2 we define our notation and discuss the similarities and differences
between 
unblended microlensing and blended microlensing 
We also give an analytic approximation that gives the underlying event
duration and peak magnification from the lensed-light fraction
and the easily measured apparent event duration and maximum magnification.

In \S~3 we discuss the usefulness of blend fits and compare with earlier work.
In \S~4 we discuss the optimal times to take follow-up data in order
	to improve recovery of parameters from the blend fit.  
In \S~5 we discuss the problem of the baseline magnitude, and
In \S~6 we discuss the problem of non-Gaussian data and whether the errors
returned by fitting programs are reliable.

\section{Degeneracies in blended lightcurves: analytic approximations for
event duration and $\Amax$}

When an isolated lens object crosses close to the line-of-sight of an isolated
background source star, the source is magnified and a microlensing lightcurve
is generated with magnification

\begin{equation}
A(u) = {u^2+2\over u (u^2+4)^{1/2}}, \ \
u^2(t) = \umin^2 + \left(t-\tmax\over\te\right)^2,
\label{eqn:A}
\end{equation}
where $u$
is the projected 
distance between the lens and source 
in units of the Einstein ring radius,
$\te$ is the time to cross the  Einstein radius, and
$t$ is time, with maximum
magnification, $\Amax$, occurring at $\tmax$.

The most important parameter is the event duration $\te$ since the
optical depth depends upon the sum of efficiency weighted event durations:
\begin{equation}
\tau = \frac{\pi}{2E} \sum_{\small\rm events} \frac{\te}
{\epsilon(\te)},
\label{eqn-tauthat}
\end{equation}
where the exposure $E$ is the product of the length in days
of the observing program and the number of observed stars, and $\epsilon$
is the efficiency of detecting an event of duration $\te$.

When other sources of light are contained in the same seeing element
as the lensed source star, the microlensing lightcurve is altered since only
a fraction of the light is actually lensed:
\begin{equation}
A'(u) = \fll A(u) - \fll + 1,
\label{eqn:Aprime}
\end{equation}
where $\fll$ is the fraction of light that is lensed (a.k.a. the blend fraction,
i.e. coming from the source star) before the lensing event 
begins.\footnote{Several terms have been used in 
different ways in the literature
for blend fraction, most commonly $f_b$ which either means the fraction
of light coming from the lens or the fraction of the light coming from
non-lens sources.  We introduce the new symbol $\fll$ to avoid the
extant confusion of nomenclature.}
Compared with unblended events,
blended photometric microlensing lightcurves suffer from a smaller maximum
magnification, $\Amaxp$ and shorter event duration $\tep$, as well
as from potential color shifts if the blended light has a different spectrum.

\begin{figure*}
\subfigure[]{\epsfig{file=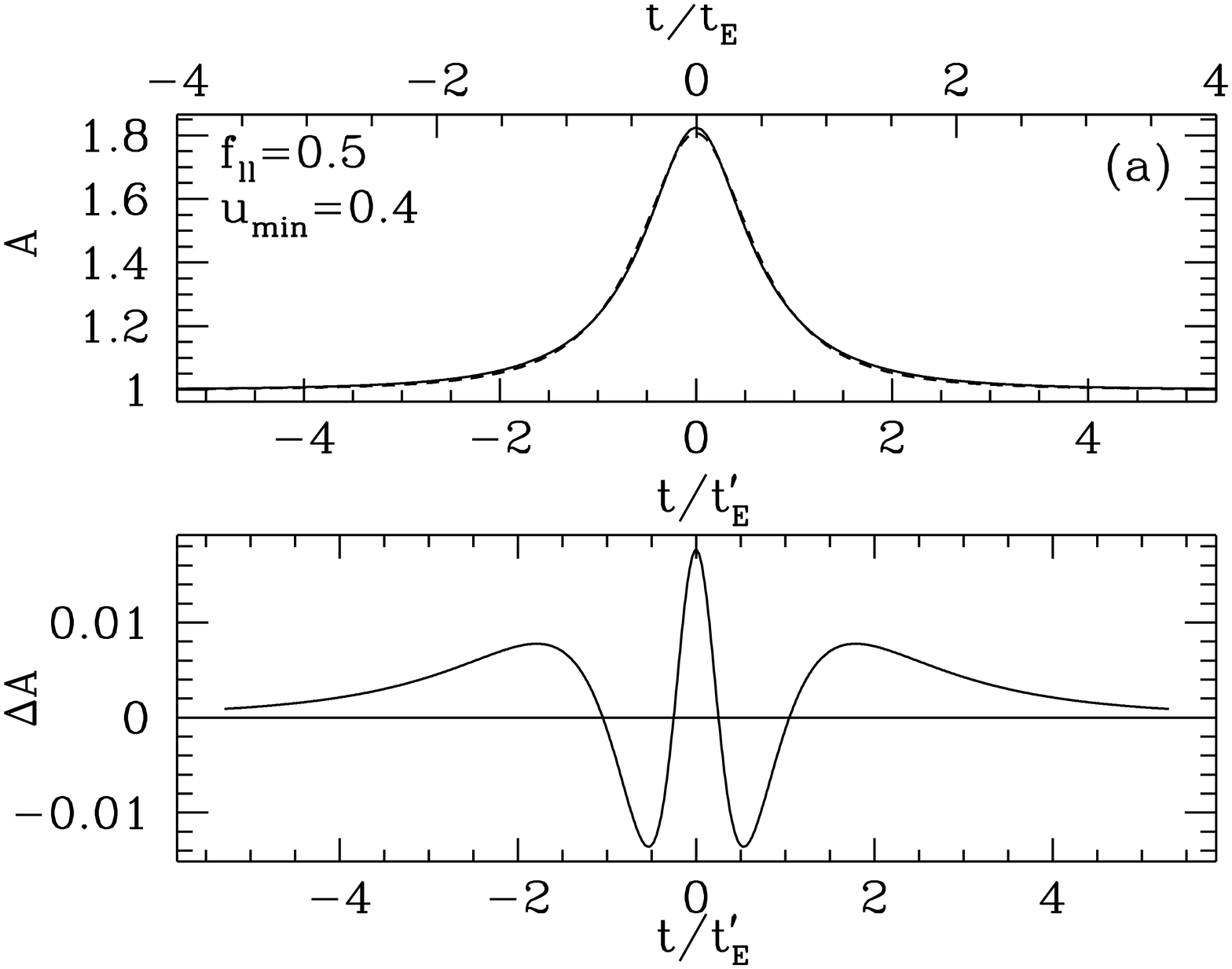,   height =.48\textwidth, angle=-90}}
\subfigure[]{\epsfig{file=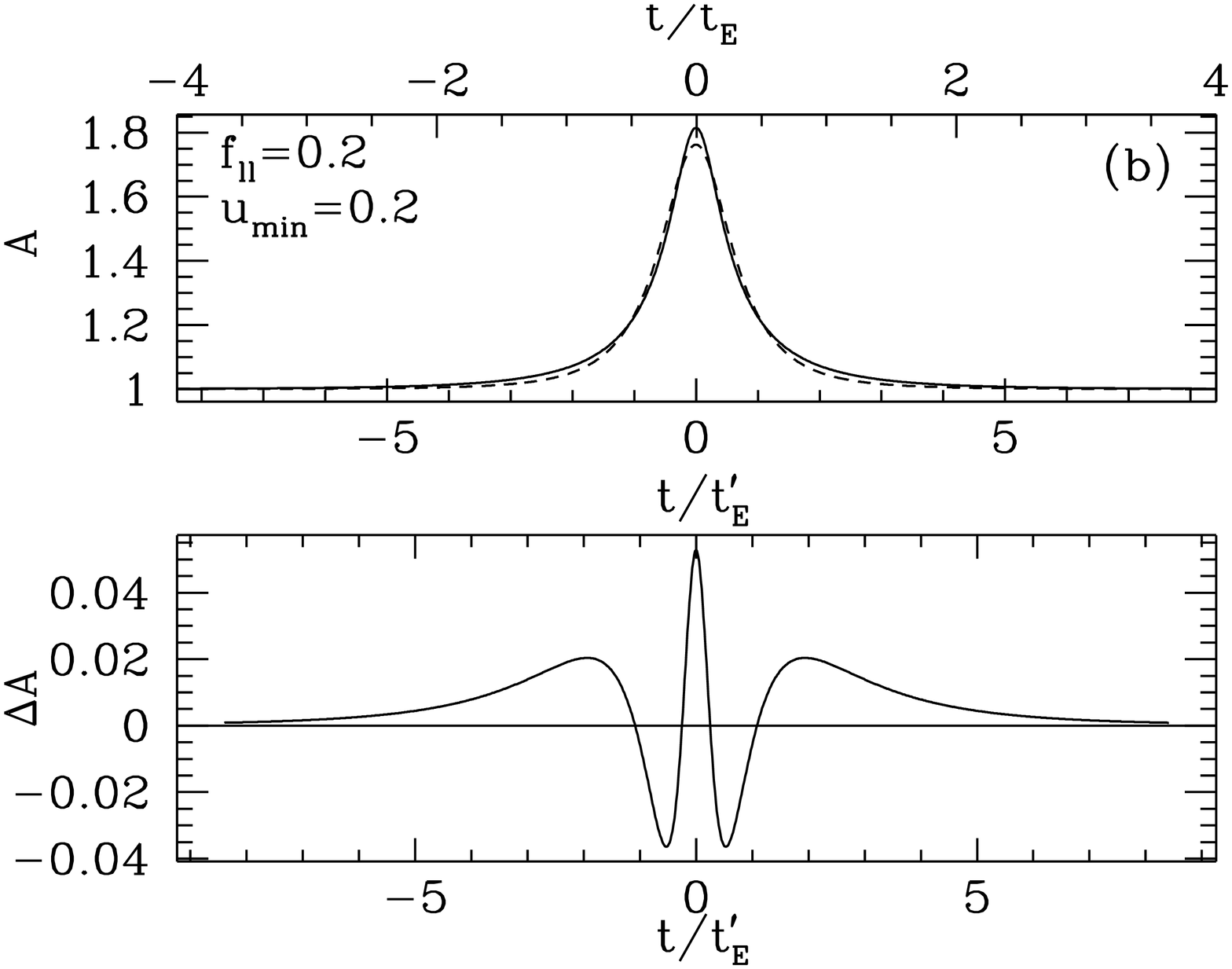,   height =.48\textwidth, angle=-90}}
\subfigure[]{\epsfig{file=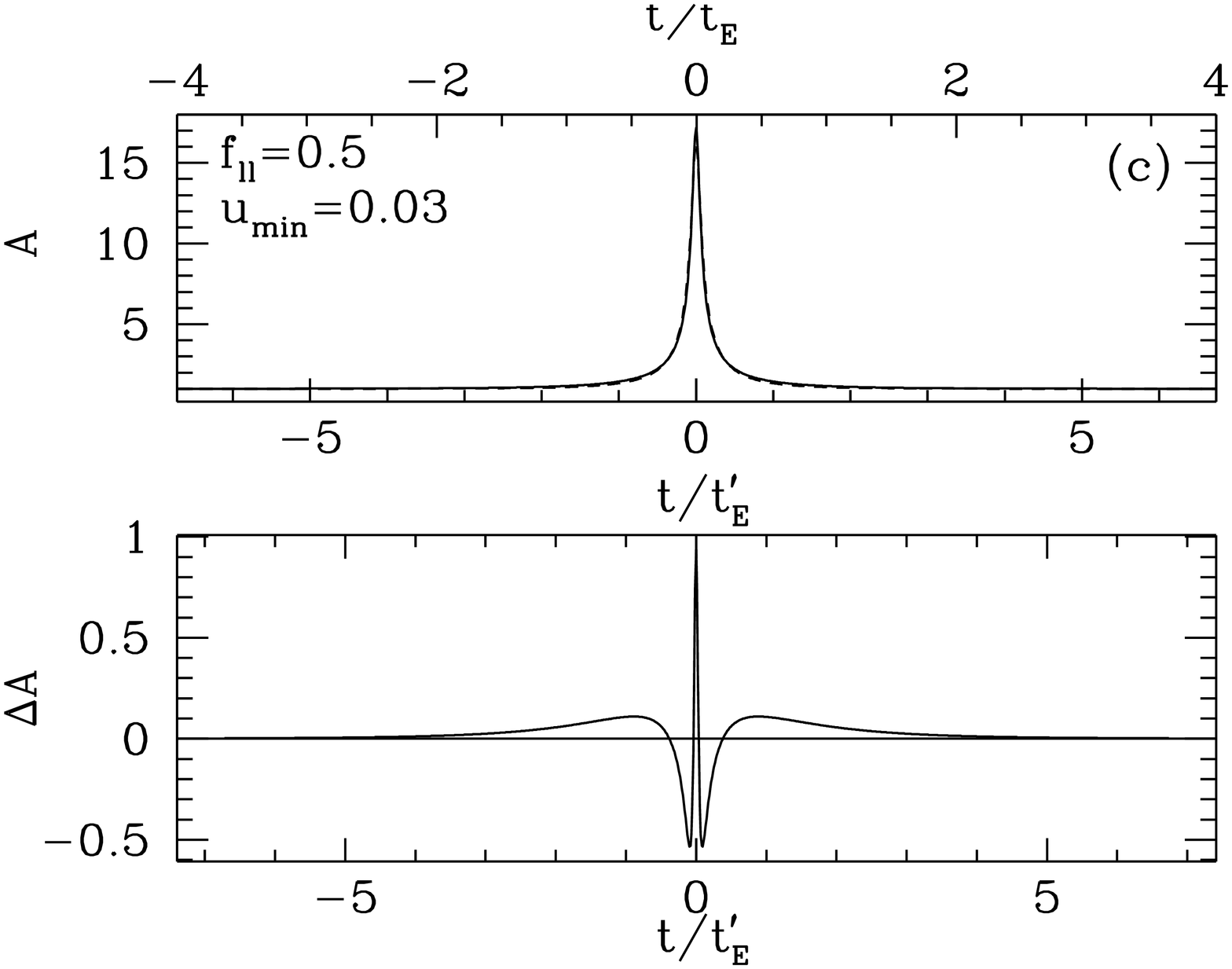,  height =.48\textwidth, angle=-90}}
\subfigure[]{\epsfig{file=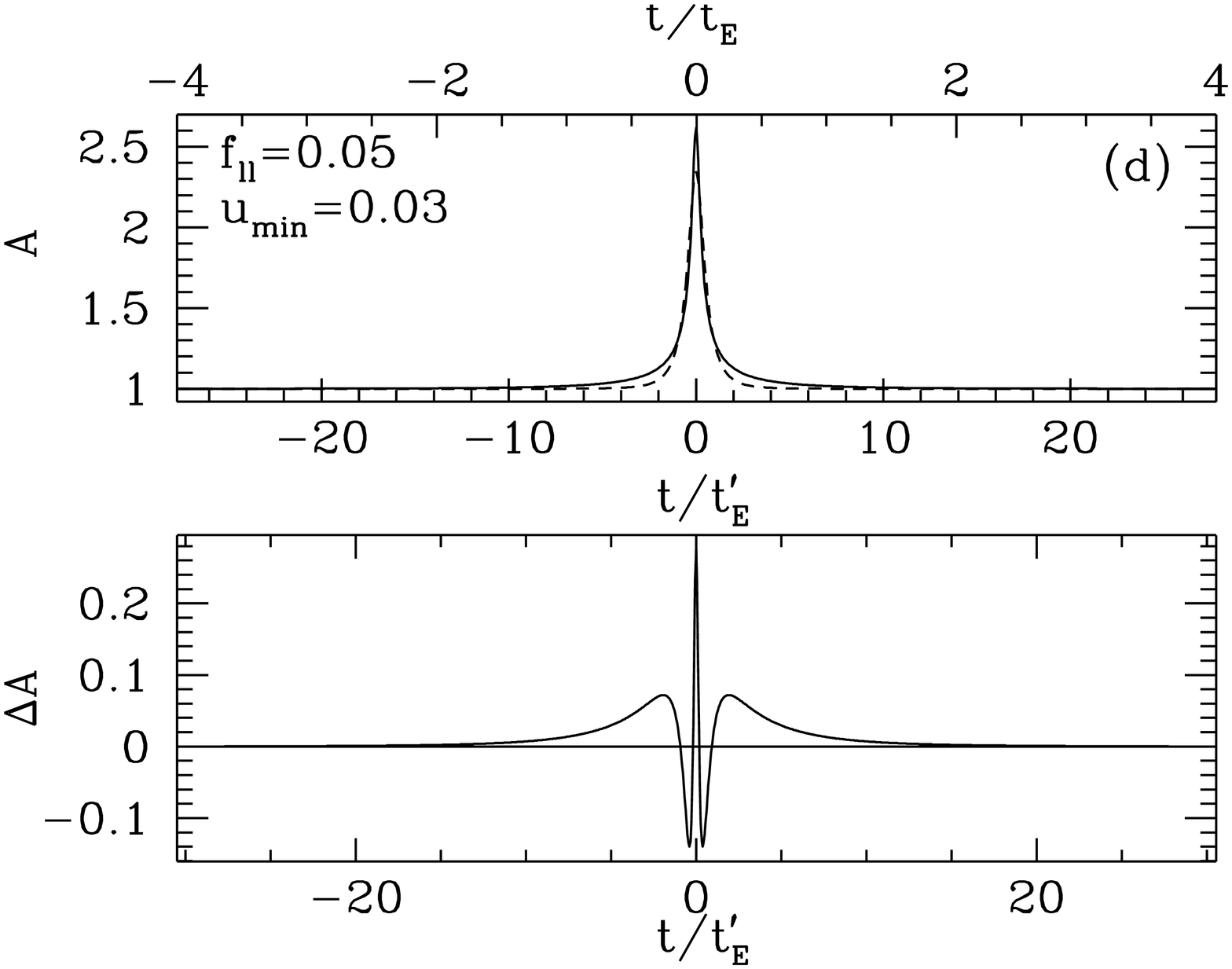, height =.48\textwidth, angle=-90}}
\caption{\label{fig:lcdiffs} Four example blended lightcurves (solid) compared with the best fit unblended lightcurve (dashed), as well as the difference, $\Delta A$, between them (blend fit minus unblended fit).  The bottom labeled time axis is in units of the {\it apparent} Einstein Ring crossing time, $\tep$, that is easily available from the data and an unblended fit. However, the extent of the time axis is $\pm 4 \te$, (labeled on the top) where $\te$ is the underlying event duration used in optical depth estimates. Thus the extent of all the time axes is roughly 160 days for a typical microlensing event of duration 20 days. Part (a) has values: $\fll=0.5$, $\umin=0.4$, $\uminp=0.634$, and $\tep/\te=0.754$.  Part (b) has values: $\fll=0.2$, $\umin=0.2$, $\uminp=0.655$, and $\tep/\te=0.475$.  Part (c) has values: $\fll=0.5$, $\umin=0.03$, $\uminp=0.062$, and $\tep/\te=0.594$.  Part (d) has values: $\fll=0.05$, $\umin=0.03$, $\uminp=0.46$, and $\tep/\te=0.144$. }
\end{figure*}

In fact, a blended lightcurve looks remarkably like an unblended lightcurve
with different values of $\te$ and $\Amax$ (e.g. Han 1999; 
DiStefano \& Esin 1995; \woz \& \pac 1997; Alard 1997).
However, as illustrated in Figure~\ref{fig:lcdiffs}, this similarity is not perfect
and there are differences in the shapes of blended and unblended lightcurves.
It is these differences which give rise to the hope that information
about blending can be extracted by fitting lightcurves with blending
parameters.  
If this similarity were
perfect, then there would be no use in fitting blended lightcurves to
photometric data.  Figure~1 shows that the shape differences are typically
small, meaning that extracting blending information will be difficult.
\woz and \pac\ 1997 (WP) 
studied this in detail and gave regions of the $\fll$, 
$\Amax$ plane where blended fits were useful and where they were not.
We return to this subject in \S~3, but the qualitative results of WP can
be seen from Figure~1, where low magnification events show a maximum difference
between the blended lightcurve and the best fit unblended lightcurve
of only 1\% or so, while the higher magnification events show more
substantial differences.

Previous workers have also given analytic formulas relating the measured
(apparent) maximum magnification $\Amaxp$, and the apparent event
duration $\tep$ to $\fll$, $\Amax$, and $\te$.
For example, \woz and \pac\ (WP) studied the degeneracies by performing
expansions of the above equations in the limits of small $\umin$ and
large $\umin$ and in these limits give the formulas relating the actual
values of $\te$ and $\Amax$ to $\fll$ and the measured $\Amaxp$ and $\tep$,
For small $\umin$ (large $\Amax$) they found
$\uminp  \approx \umin/\fll$,
and $\tep  \approx \fll \te $, while in the limit of large $\umin$ 
($\Amax \approx 1$) they found $\uminp \approx \umin / \fll^{1/4}$,
and $\tep \approx \fll^{1/4} \te$. 

DiStefano \& Esin (1995), and Han (1999), and Alard (1997) 
took a different approach, 
solving equation~\ref{eqn:Aprime} 
for $\Amax$ and giving the actual $\te$ in terms of $\tep$ by
requiring that the two different parameterizations give the same amount 
of time with $A>1.3416$.  They found: 
\begin{equation}
{\Amax}_{(\rm HDE)} = (\Amaxp - 1 + \fll)/\fll,\ \ \
{\te}_{(\rm HDE)} = \tep \left( {u_1^2 -\umin^2 \over 1-{\uminp}^2} \right)^{1/2},
\label{eqn:Ahan}
\end{equation}
where $\uminp= u(\Amaxp)$, and $u_1 = u(A(A'=1.3416))$, can be found from
the inverse of equation~\ref{eqn:A}: 
\begin{equation}
\label{eqn:u_of_a}
u(A) = (2/\sqrt{1 - 1/A^2} -2)^{1/2}
\end{equation}

Noting in Figure~1 that the differences between the blended and unblended
lightcurves tend to be large in the peak, and that the 
values of $\tep$ and $\Amaxp$ 
are found by fitting, we worried that the HDE formula, which assumes
equality in the peak, might not be accurate.  We also wondered about
the range of applicability of the WP formulas and so decided to
test these formulas.
We did this by fitting artificial blended lightcurves with an unblended 
source model and finding the best fit values of $\tep$ and $\Amaxp$.
We also fit these lightcurves with blended source models and correctly
extracted the input blend parameters.

As shown in Figure~2,
we found that the WP formulas are not very useful over most of the parameter
range, and that the HDE equations work well only over a restricted range of
parameters.  For the WP formulas this is not surprising since they were created
only to show that the degeneracies exist in certain limits.
For relatively large lensed-light fraction
and for relatively low values of $\Amax$ the HDE equations give a good
estimation of the best fit $\Amaxp$ and $\tep$, but for small 
lensed-light fraction or high $\Amax$ the estimates of these equation can 
be far off.  

As expected, it is just where the blended lightcurve shape differs the
most from an unblended fit that the HDE approximations
do not work well.  
The reason can be seen in Figure~1, where for high magnification events
and low lensed-light fraction the blended lightcurve differs strongly in the
peak area, but not so much in the lightcurve middle rising and falling
regions.  Thus, the best fit
unblended lightcurve will allow the actual peak magnification to
overshoot and compensate for these points by undershooting in the middle 
regions.
Since the HDE formula forces the lightcurves to match at the peak and
when $\Amaxp=1.34$, it will overestimate the best fit peak magnitude
and underestimate
the event duration. 

By studying many such examples, one can come up with a formula that 
does a better job of relating the best fit $\Amaxp$ and $\tep$ to
$\Amax$, $\te$, and $\fll$ in
the parameter ranges where the HDE formula does not work well.
The points in Figure~2 show the best fit values of $\Amaxp$ and $\tep$ 
vs $\fll$ found 
by fitting artificial blended lightcurves.  The dashed lines show the HDE
estimates and long dash lines the WP estimates for $\tep/\te$.
At small values of $\Amax$ 
($<3$) the HDE formulas do work very well (better than the new formula)
and they should be used.
However for $\Amax>3$ the HDE formulas do not give accurate estimates.
To find a better approximation, 
one can
fit a straight line to the data for a given $\umin$ and get a 
formula which fits well except for very low lensed-light fraction.  
Repeating this procedure for
different values of $\umin$, one discovers
that the slopes and zero points of the linear fits are also quite linear in
$\umin$.  Thus a simple fitting linear formula that covers much 
of the parameter space can be found.  However, if one fits a quadratic
for the low $\fll$ events one can get an even better formula which works 
very well for $\fll<0.3$.
Thus we find an approximation: 
\begin{eqnarray}
  \Amax &\approx& 
  \begin{cases} 
    A_{\rm HDE}, 	&\text{if} \Amax < 3;\cr
    {\Amaxp - 0.9785 + 0.4150 \fll \over 0.8153 \fll + 0.00021}, 
    &\text{if} \Amax>3 {\rm\  and\ } \Amaxp<10;\cr
    {\Amaxp - 0.3618 + 0.2106 \fll \over 1.0282 \fll - 0.04433}, 
    &\text{if} \Amax>3 {\rm\  and\ } \Amaxp > 10,  ;\cr
    \end{cases}
    \nonumber\\
    \tep/\te &\approx& 
    \begin{cases} 
	{\tep}_{(\rm HDE)}/\te, 
	&{\rm if}  \Amax<3; \cr
	(-1.0946 \umin +0.9418)\fll + 1.141\umin + 0.0564, 
	&{\rm if} \Amax>3 {\rm \ and\ }\fll>0.3; \cr
	\boldsymbol{F C U},
	&{\rm if} \Amax>3 {\rm \ and\ }\fll<0.3 , \cr
  \end{cases}
  \label{eqn:quadfitnum}
\end{eqnarray}
where
\begin{eqnarray}
  \boldsymbol{F} = \left(1, \fll, \fll^2\right),\ \ 
  \boldsymbol{U} = \left(
  \begin{matrix}
    1 \cr
    u_{min} \cr
    u_{min}^2\cr
  \end{matrix}
  \right),
\end{eqnarray}
\begin{equation}
  \boldsymbol{C} = \left(
  \begin{matrix}
    0.02548 & 1.0626 & -1.6504 \cr
    1.1914 & 7.284 & -11.50 \cr
    -0.8824 & -26.58 & 44.28 \cr
  \end{matrix}
  \right),
  \label{eqn:cmatrix}
\end{equation}
and $\umin$ is found from $\Amax$ and equation \ref{eqn:u_of_a}.

In using this formula, one typically starts with measured values
of $\tep$, $\Amaxp$, and an initial guess of $\Amax$ and uses different (unknown) values of $\fll$,
to find the corresponding underlying $\Amax$ and $\te$.  If the value
of $\Amax$ found using the new fitting formula is smaller than 3, then
one should use the HDE formula instead.

The new fitting formula is shown as the solid line in Figure~2
and does better than HDE or WP for $\Amax>3$.
Over the range $0.01 < \fll < 1.1$, and $3< \Amax< 70$ the new fit formula
gives a typical error in $\te$ (compared with actually fitting
the microlensing lightcurve with a blend fit model) of
around 3\% and a maximum error of 9\%.  For $\Amax$ the typical error
is 4\% and the maximum error is 12\%.  The HDE formula can be off by
more than 50\% in $\te$ and 24\% in $\Amax$ in this region of parameter space.

\begin{figure*}
\subfigure[]{\epsfig{file=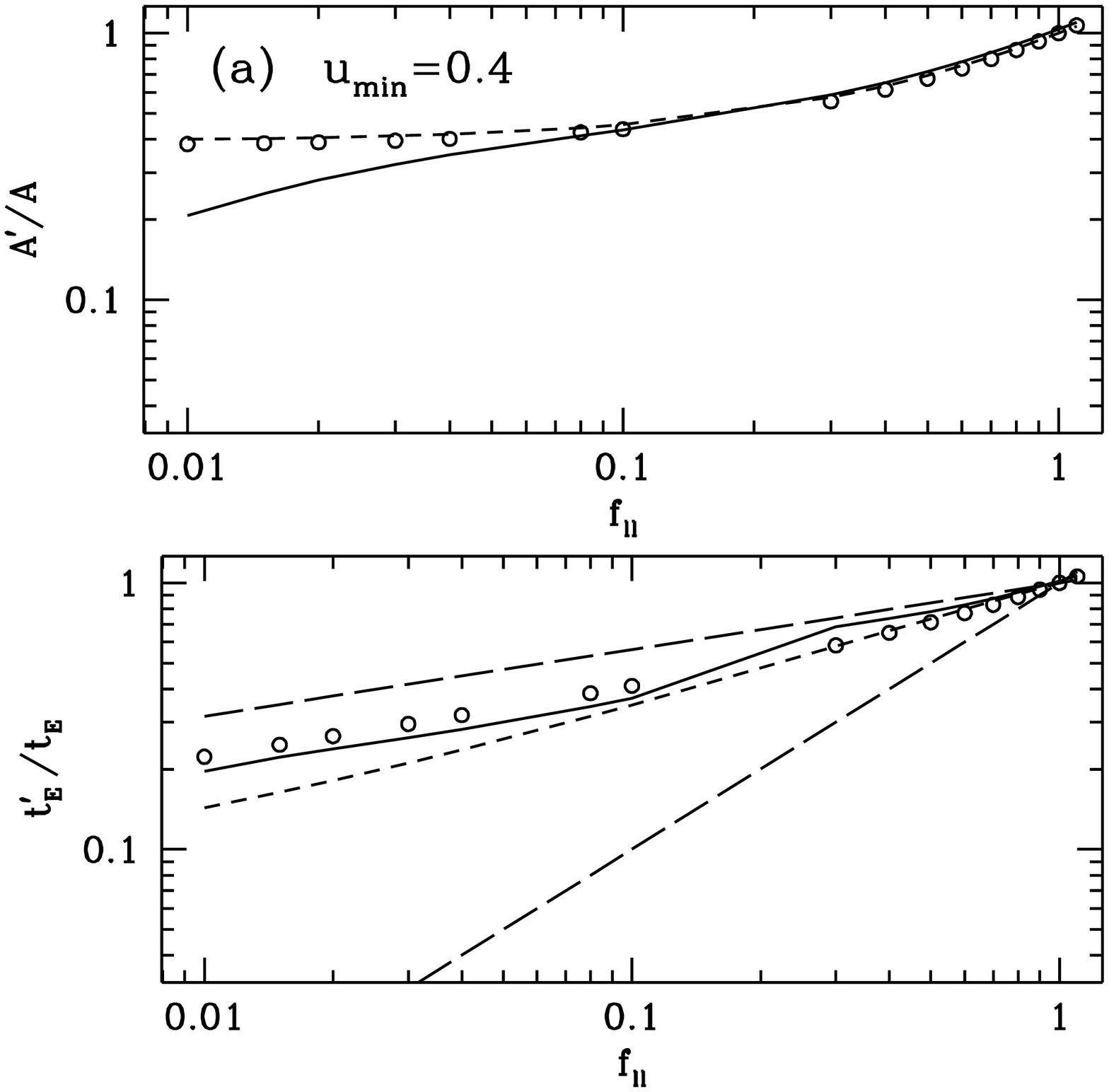, width=.48\textwidth}}
\subfigure[]{\epsfig{file=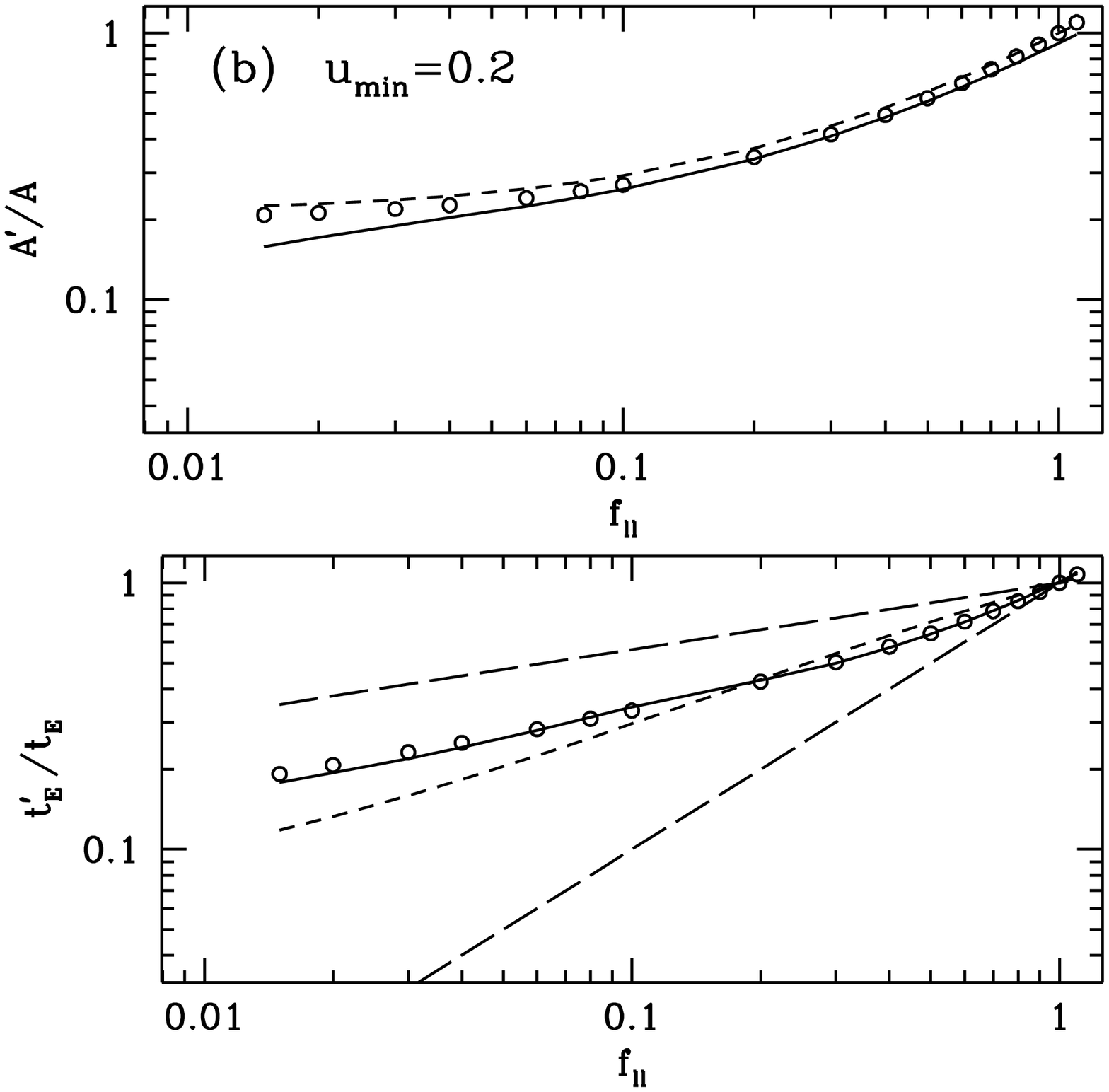, width=.48\textwidth}}
\subfigure[]{\epsfig{file=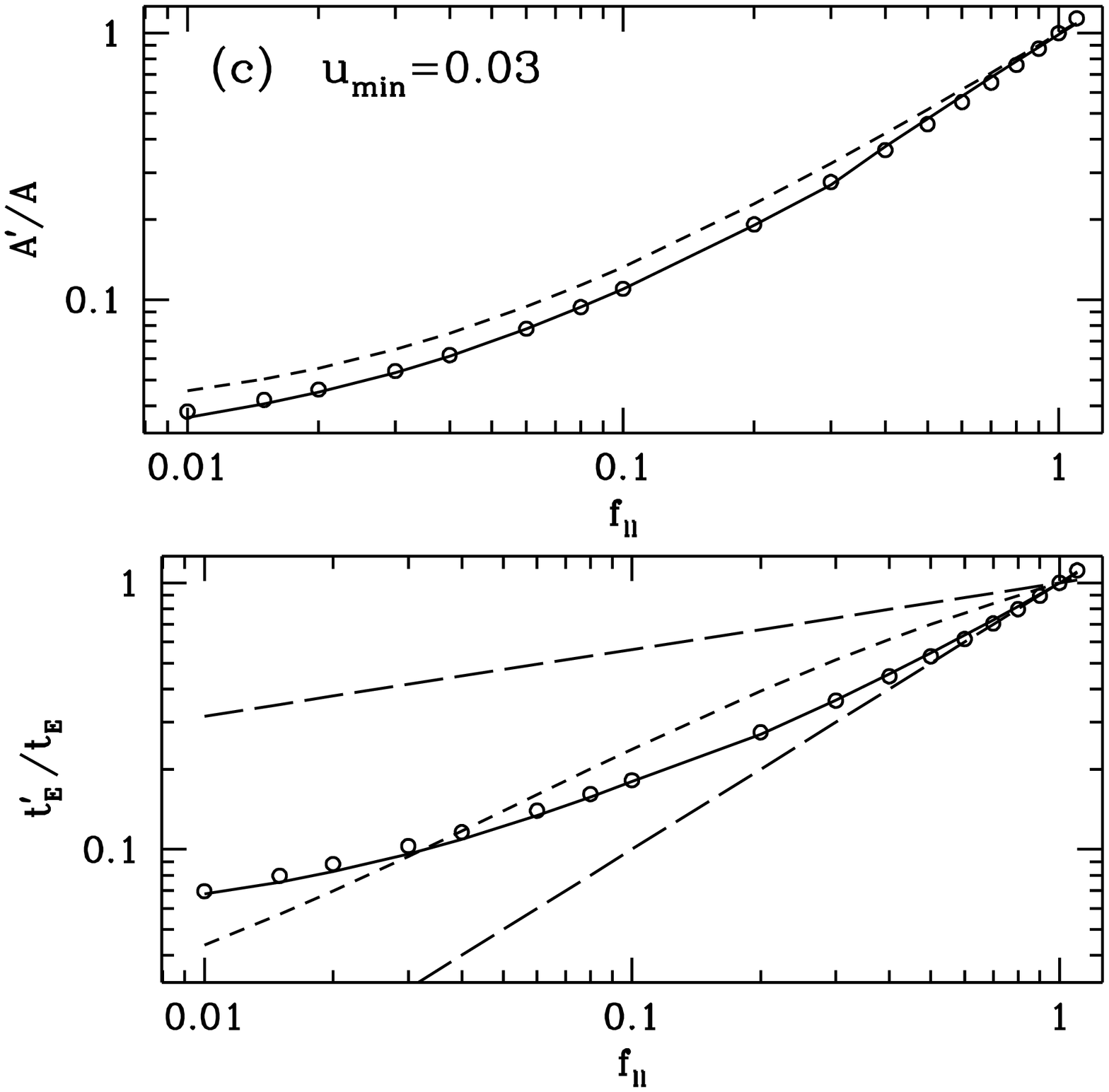, width=.48\textwidth}}
\caption{Comparison of approximation formulas for relating the underlying microlensing event duration $\te$ and maximum magnification $\Amax$, to the blend fraction, $\fll$, and easily measured apparent event duration and maximum magnification, $\tep$, and $\Amaxp$.  The circles give the actual blended and unblended results from our lightcurve fitting program, while the solid lines show our new approximation formula.  The short dashed line shows the HDE approximation, while the long dashed lines show the WP approximations in their two limits.  Part (a) is for an actual $\umin=.2$ ($\Amax=5.07)$, part (b) shows $\umin=.03$ ($\Amax=33$), and part (c) shows $\umin=.4$ ($\Amax=2.65$).}
\end{figure*}

In summary,
we tested the HDE formula, \woz and \pac\ (WP) formulas 
and Equation~\ref{eqn:quadfitnum} over a wide range
of parameters and found the new fitting function works better 
than HDE for all values of $\fll$
when $\Amax>3$ and $\Amaxp>1.34$, while the old HDE formula works better
for low values of $\Amax$ and $\Amaxp$.  The WP large $\Amax$ formula gives
$\te$ within 10\% only for large $\fll (>0.5$), and large $\Amax$,
while the other WP formula is not useful except for $\Amax \ll 1.34$.

Since in microlensing experiments the event durations are found
by photometric fitting and since
the optical depth is proportional to the sum of the fit $\te$'s,
when making corrections for blending 
it is important to properly relate the lens-light fraction of each event
to the underlying event duration.

\section{Usefulness of blend fits}
\woz and \pac\ (1997) (WP) studied the degeneracy of blend 
fits and concluded that in many cases
blended and unblended lightcurves cannot be distinguished by photometric
fitting.  They described areas of parameter space where blend fits
would be useful and areas where they would not.
While we think that WP did an accurate and very useful
calculation, and we agree with their conclusion that blend fits are usually
not very useful, we wanted 
to repeat their analysis for several reasons.  First,
WP did not include the baseline magnitude in their fits, reasoning that
since many measurements are taken before and after the event, the error
in baseline magnitude was not significant.  In fact, we find that
error in the baseline magnitude is one of the most severe problems
in blend fits.  We find that errors even at the few percent level can
drastically alter the parameter values extracted from the fit.
Second, WP considered only evenly spaced observations and
we wanted to consider whether different follow-up strategies
could improve the ability to extract the parameters.

In our studies,
we find the error in fit parameters three ways.  First we create
artificial lightcurves using the theoretical formula and add
Gaussian random noise to each measurement. 
We perform blended and unblended fits on these lightcurves using Minuit (CERN Lib. 2003).
Second we calculate the error matrix by inverting the Hessian matrix
as discussed in Gould (2003). Finally to understand the effect of the non-Gaussianity of the errors
in real microlensing experiments we create artificial lensing lightcurves
by adding microlensing signal into actual non-microlensing 
lightcurves obtained by the MACHO collaboration,
and then fit these.

\begin{figure}[t]
\epsfig{file=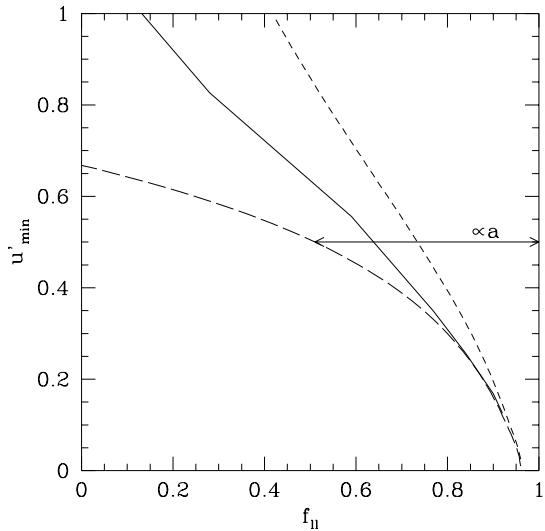,width=.48\textwidth}
\caption{Comparison of our results to the previous results of \woz and \pac one sigma limits on $f_{ll}$ for the range of apparent $u_{min}$ (from non-blend fits).  The solid line is from WP, the long dashed line is the $1-\sigma_{f_{ll}}$ limit for a non-blend fit, and the short dashed line is the value of $f_{ll}$ for which  $f_{ll}+\sigma_{f_{ll}}=1$ for a blend fit.  In the region below the long dashed line blending is detectable at the one sigma level, above the short dashed line blended events are indistinguishable from unblended events, and inbetween the two dashed lines detection is marginal.  The region where blending is distinguishable can be scaled with $a$ (eqn. \ref{eqn:WPa}). \label{fig:wp}} 
\end{figure}

Since the method of calculating the error matrix is closest to what WP did, we first give these results.  Briefly, we calculate the Hessian matrix (the matrix of second derivatives of the light curve residuals with respect to each parameter) then invert it.  The square root of the diagonal elements of the resulting matrix are then the one sigma errorbars of the parameters.  This accounts for correlations in the parameters, but not any nonlinearities.  WP used a very similar method, but used it to calculate the $\Delta \chi^2$ instead of the error bars. 
In figure \ref{fig:wp} we show that our method brackets WP's.  
We show limits calculated as both the one sigma lower limit on $f_{ll}$ for an unblended lightcurve and the value of $f_{ll}$ that gives $f_{ll}+\sigma_{f_{ll}} = 1$.

We note, as WP found, that parameter errors scale linearly with 
\begin{equation}
a = \frac{\sigma}{\sqrt(N)} 
  \label{eqn:WPa}
\end{equation}
for $N$ points taken during the peak (defined as lasting $4 \te$)\footnote{This is true for large enough value of $N$, for small values of  $N \lesssim 16$ parameter errors increase faster than $a$.}.  Thus our results can be scaled for other numbers of observations with different values of $\sigma$.
Thus, we find that our results agree with those of WP if we assume the
baseline magnitude is known and take a uniform sampling.

\section{Follow-up observations}

Figure~1 shows that the difference between blended and unblended lightcurves
is not always uniform across the lightcurve.  So if one wanted to 
plan follow-up observations to improve the accuracy of the blend fit,
one should concentrate on the regions of the lightcurve where the
differences are largest.  Thus it may be possible to do better
than WP suggested with their equal spaced observation calculations.
To test this hypothesis we calculated the error matrix for blended fits adding in follow-up observations at different points on the lightcurve.  
As seen in the Figure~1 examples, 
for any choice of parameters there are five places where 
the difference lightcurves are maximum, and therefore where
follow-up data is more useful than average: at the peak,
in the rising/falling portion of the curve, and in the wings.  
The precise locations change with the choice of parameters but 
for Figure 1a they are found to be
localized near the peak at $(|t/t_E|<0.1)$, in 
the falling (or rising region) at $(0.3<|t/t_E|<0.6)$, and 
near the baseline at $(1.0<|t/t_E|<1.5)$. 
Observations taken
between these regions 
do little in constraining the parameters.  
In addition points $t$ greater than $2 t_{E}$ are very helpful because they 
fix the baseline in our simulated lightcurve.
We discuss the baseline
separately in \S~\ref{sec:baseline}.   
\begin{figure}[t]
\epsfig{file=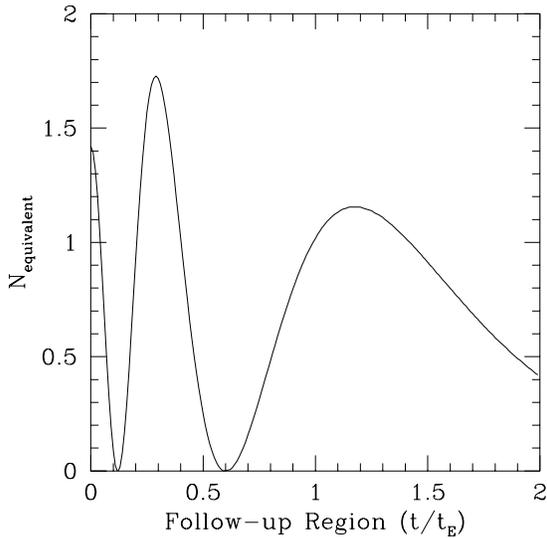,width=.48\textwidth}
\caption{The equivalent number of uniform follow-up data points required to improve measurement of $f_{ll}$ as much as a single follow-up observation is plotted as a function of when the single follow-up observation is taken.  In this case $u_{min} = 0.25$, $f_{ll}=0.25$, and 4 follow-up points are added. 
Times with $N_{\rm equivalent}>1$ are the most effective, while times with
$N_{\rm equivalent}<1$ are less useful.
\label{fig:fur}} 
\end{figure}
In Figure~\ref{fig:fur} we compare the relative value of added points as a function of the time they are added.  We find that, in this case, with 40 observations, 4 extra focused observations can reduce the error on $f_{ll}$ by 7.7\%.  To get the same reduction of error on $f_{ll}$ with evenly distributed observations we would need 7 observations, in other words, each added focused observation is equivalent to increasing the sampling by 1.75 points over 4 $t_E$.  
The numerical value of the extra effectiveness obtained using focused 
versus evenly spaced
observations varies with underlying parameters and 
the total number of added points.   

Precise follow-up measurements at multiple focused locations can 
improve the determination of $f_{ll}$ even more as they further constrain the shape of the lightcurve.  
\begin{figure}[t]
  \epsfig{file=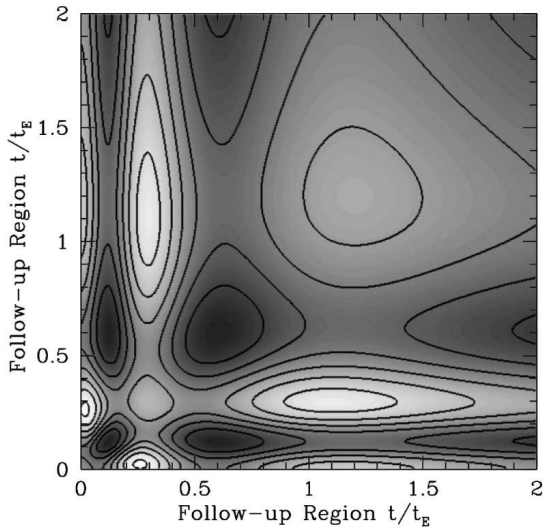, width=.48\textwidth}
  \caption{Number of additional uniformly distributed observations required for the same improvement as 8 focused observations (2 follow-up regions each with 4 observations).  The seven contours are 2 (darkest regions), 4, 6, 8, 10, 12, \& 14 (lightest regions).  In this case $u_{min}=0.25$ and $f_{ll}=.25$.  Values above (below) 8 indicate an advantage (disadvantage) relative to uniform follow-up.   \label{fig:contour}}
\end{figure}  In order to see the effect of adding multiple follow-up observations at two distinct times we compare this with adding evenly spaced observations.
In Figure~\ref{fig:contour} we plot the increase (or decrease) in 
effectiveness of extra focused observations as a function of the two times at
which they are taken.  The contours around the light areas show regions
of increased effectiveness, while dark areas show areas where the focused
observations are less valuable than evenly spaced observations. In this
example the effectiveness is increased by up to a factor two.

 It is important to note that with more observations or higher accuracy in each follow-up region the advantage per added observation is reduced and the relative values of the various minimums vary, though they stay in roughly the same place.  For practical use it is important to note that the time of the optimum second follow-up observation(s) varies with the time of the first follow-up observations(s).  In practice one would need to calculate optimum observing times for an event in progress as a function of all the previous measurements.  
 
\begin{figure}[t]
  \epsfig{file=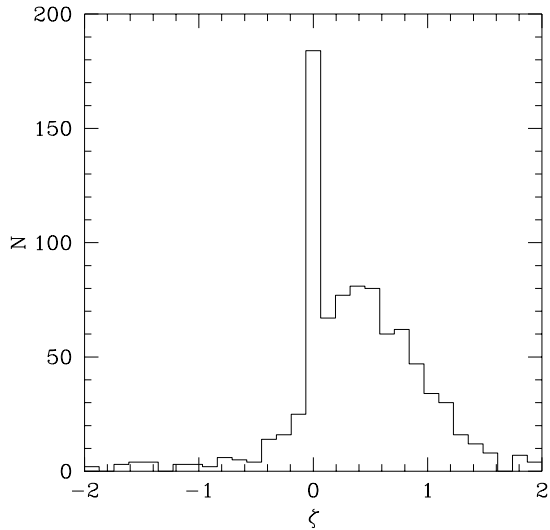, width=.48\textwidth}
  \caption{The ratio $\zeta$ showing the advantage of a focused follow-up strategy for 71\% of events ($\zeta < 1$ - unshaded region).  For 15\% of the events in our simulation are outside the range of this plot $|\zeta| > 2$. \label{fig:furdiff}}

\end{figure}

One problem with the above approach is that without knowing the underlying parameters, particularly $t_E$, it is difficult to predict the best times to take follow-up data.  To test if a practical experiment could be designed to take advantage of focused follow-up data we simulated an experiment.  First we generated lightcurves with 80 points over $8t_E$ with .05 Gaussian errors at the baseline drawing $f_{ll}$ randomly from the interval $[.01,1)$ and $u_{min}$ randomly from the interval $(0,1)$ requiring $A_{max}^\prime > 1.34$ in the HDE approximation.  We also adjusted $t_E$ to keep $t_E^\prime \sim 10{\rm days}$ also using the HDE approximation.  We then generated 9 follow-up observations over 3 days at the peak and fit the first half of the light curve plus the follow-up data.  From this first fit we calculated the optimum times for two more bouts of follow-up. We generated these, both with 9 observations over 3 days, and then fit the entire lightcurve with the added 27 points.  We also generated 27 points of follow-up uniformly distributed over the 20 days starting at the peak, added it to the initial lightcurve and fit the resulting data.  
To see the relative improvement for the two methods we calculate a parameter $\zeta = (f_{ll_{focused}}^\prime-f_{ll})/(f_{ll_{unfocused}}^\prime-f_{ll})$,
which is the ratio of the error in blend fraction
given by focused observations to the error in blend fraction
given by uniform follow-up sampling.
We plot the distribution of $\zeta$ in Figure~\ref{fig:furdiff} finding
that our strategy gives an improvement $(\zeta < 1)$ for 71\% of the 
events and a worsening in 29\% of the events.
We find a substantial improvement ($\zeta < .5$) for 45\% of
the events, and and even larger improvement ($\zeta < 0.1$) 18\% of
the time.  
Thus we conclude that for the the same amount of observing time we 
can make a more accurate measurement of $f_{ll}$ by focussing the follow-up
observations.

In summary we find that observations concentrated at a few times can constrain the microlensing parameters as well as many measurements distributed throughout an event.  The best place for these measurements are at the peak, 
in the falling/rising portion, and in the wings with regions between where added observations do no good.  In most cases it is possible to constrain the event parameters well enough with the first half of the data and some follow-up observations near the peak to predict the last two optimum observing times.

\section{Baseline Magnitude \label{sec:baseline}}
\begin{figure}[t] 
\epsfig{file=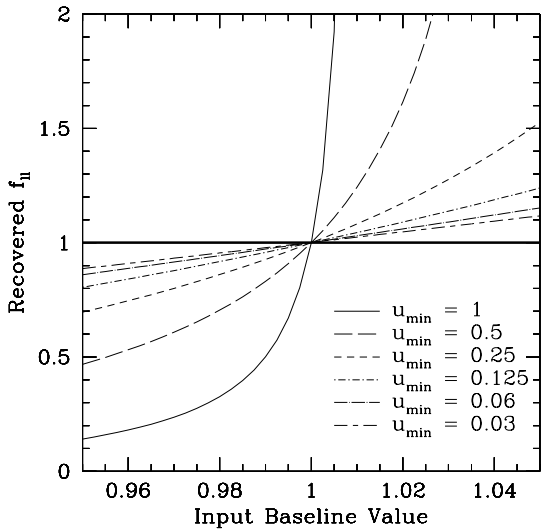, width=.48\textwidth}
\caption{\label{fig:baseline} The recovered $f_{ll}$ for unblended light curves as a function of an input baseline magnitude (fixed at a given value).  Forty
points over 4 $t_E$ are used.}
\end{figure}

The baseline magnitude of a lightcurve can in principle 
be very well determined since
many measurements can be taken before or after the microlensing event.
WP assumed that this was the case and so did not include the baseline
magnitude as one of their fit parameters.  In real microlensing surveys,
however, it may be that the error in average magnitude is not entirely
statistical, and may not average down as expected.
There may be a systematic limit to the accuracy with which the baseline magnitude can be determined. In fact, detectors and telescope systems drift over time and so measurements made much later may actually reduce the accuracy of the baseline magnitude.  To investigate the importance of the baseline magnitude, we created artificial lightcurves without any errors and fit them with a model with a fixed value of baseline magnitude that differed from the actual baseline magnitude by various amounts.  Our results are shown in Figure~\ref{fig:baseline}.  We find that the dependence on baseline is very strong for low amplification events and not as strong for higher amplifications events, but in
any case even a 2\% error in baseline magnitude determination can strongly
bias the recovered blend fraction.

\begin{figure}[t] 
\epsfig{file=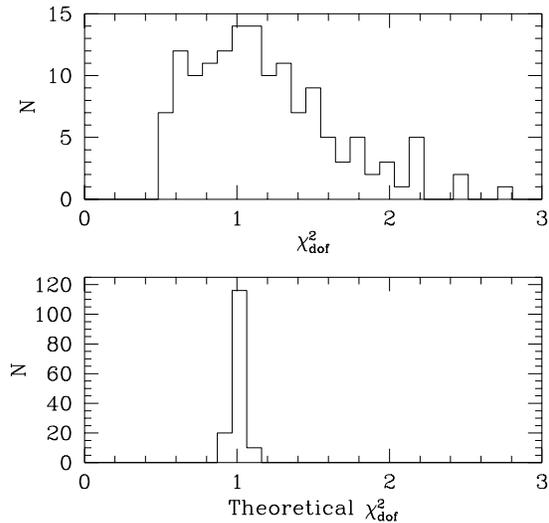, width=.48\textwidth}
\caption{Actual distribution of $\chi^2/n_{dof}$ (top) for MACHO data and theoretical distribution of $\chi^2/n_{dof}$ (bottom) \label{fig:blchisq}}
\end{figure}

Next, to see how well baseline magnitudes converge in real data, we used the
MACHO collaboration database of random stars (Alcock, \etal 2000).  
We looked at the $\chi^2/n_{dof}$ of a fit to a constant lightcurve
for our real lightcurves and compared this to simulated
ideal lightcurves with the same number of points and Gaussian errors.  
For the simulated Gaussian lightcurves we find the $\chi^2/n_{dof}$ 
distribution peaked near unity and distributed as expected, but 
for the real data the distribution of $\chi^2/n_{dof}$ is much broader. 
These two distributions are shown in Figure \ref{fig:blchisq}.

As an estimate of the error in the baseline 
which arises due to the systematic drift and non-Gaussian nature of the 
magnitude errors, we calculated mean and median for the points in 
each of 146 lightcurves in MACHO field 119, one of the most frequently observed fields. 
We found that the distribution of mean minus median had a dispersion of 1.3\% indicating that the error in the baseline flux is $\sim1.3\%$.  
Referring to Figure~\ref{fig:baseline} we see that for a typical event with $u_{min}=0.5$,
this implies a typical spread in $f_{ll}$ of 0.18 due to baseline alone.  Since half of all events
have $u_{min}<0.5$, half of all events will have an even larger bias.
For more sparsely sampled fields this dispersion due to error 
in baseline fit would be even larger.

\section{Errors in Fit Parameters}
From Macho Project data (Table 6 of Popowski et al.\ 2005) it seems blend fits return biased parameters.  For the set of Macho clump giant events, which are believed to be minimally blended from their positions on the color magnitude diagram, many are best fit with blending.  If the events are not blended then a systematic bias in the fits must make them appear to be blended.  A systematic bias in recovered lensed-light fraction would lead to a bias in the optical depth as well.  The MACHO collaboration investigated the blending of their clump giant sample and decided to use the parameters from the unblended fits.  They also used a subsample of events that were less likely to be blended to check for a bias due to blending and found no such bias.

\begin{figure}[t]
\epsfig{file=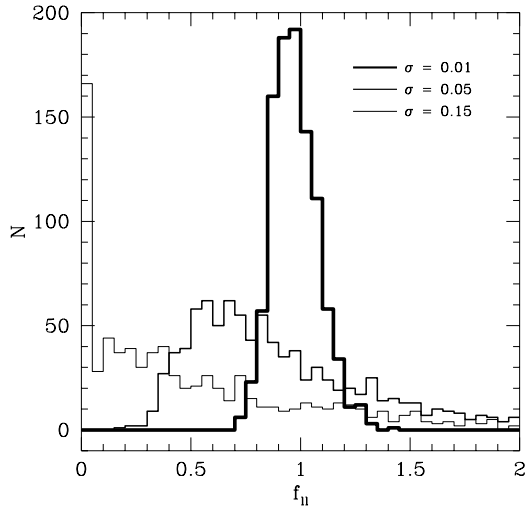, width=.46\textwidth} 
\caption{\label{fig:fllhists} Recovered $f_{ll}$ for data with Gaussian errors of $0.01, 0.05,$ and $0.15$.  As the errors on individual data points increase the distribution becomes increasingly skewed with the mode shifting toward 0 for larger errors.  Also note that 11\% of the $\sigma = 0.05$ events and 24\% of the $\sigma = 0.15$ events had recovered $f_{ll}^\prime > 2$ while none of the events with  $\sigma = 0.01$ were fit best with $f_{ll}^\prime > 2$.}
\end{figure}

To test for a systematic bias we generate 1000 lightcurves with Gaussian errors
for each of three different values for the error on each point: $\sigma = 0.01$, $\sigma = 0.05$, and $\sigma = 0.15$.  The recovered lensed-light fractions for these events are shown in Figure \ref{fig:fllhists}.  As the error on individual datum increases the distribution of $f_{ll}^\prime$ becomes increasingly skewed.  
We find that while the mean $f_{ll}^\prime$ may not decrease, the most 
probable value does decrease.  
This reduction in the mode is at least partially compensated by the large tail of the distribution with $f_{ll}^\prime > 1$, but for the small number of events a microlensing experiment 
observes it is unlikely that many of the few events with $f_{ll}^\prime \gg 1$ will be observed.  Even if one event with $f_{ll}^\prime \gg 1$ is observed it may be ignored as it is an unphysical value of the parameter, thus leading to an underestimate of the average value of $f_{ll}$. 
Thus we find that as the errors in measurement increase blend fitting 
becomes more and more likely to return biased results.  The direction of
the bias is more often toward small values of $f_{ll}$.  Thus events
that are in reality unblended become more and more likely to return fit
values implying that they are heavily blended.

\section{Conclusions}
We find agreement with previous workers that blend fits are problematic, but
can be useful especially for high magnification events.  When performing
blend fits it is helpful to get extra measurements near the peak and at
other specific points along the lightcurve.  We find that if care is not
taken in the treatment of the lightcurve baseline magnitude the fit
results can be severely biased and in real data the
errors returned on fit parameters should be treated with caution.  We find that
blend fits return a biased, skewed distribution of the underlying parameters tending to indicate more blending than actually exists.  
Finally, note that when the microlensing
event contains signal from other physical effects such as weak parallax
or binary effects blend fits can yield unreliable results.  
These effects are not rare, and since the difference
between blended and unblended lightcurves is small, even an almost
undetectable real deviation
from the standard point-source-point-lens lightcurve can render blend fit
results meaningless.

\acknowledgments
We thank David P. Bennett and Piotr Popowski for many useful 
discussions on the topic of blending.
This work is supported in part by the DoE under grant
DEFG0390ER40546.

\end{document}